\documentstyle[12pt,epsf]{article}
\begin{document}
\begin{center}
{\large {\bf Non-perturbative regularization and renormalization: simple 
examples from  non-relativistic quantum mechanics}}\\
\vspace{12pt}
Daniel R. Phillips, Silas R. Beane, and Thomas D. Cohen\\
\vspace{12pt}
Department of Physics, University of Maryland, College Park, MD, 
20742-4111
\end{center}

\begin{abstract}
We examine several zero-range potentials in non-relativistic quantum
mechanics.  The study of such potentials requires regularization and
renormalization.  We contrast physical results obtained using
dimensional regularization and cutoff schemes and show explicitly that
in certain cases dimensional regularization fails to reproduce the
results obtained using cutoff regularization.  First we consider a
delta-function potential in arbitrary space dimensions.  Using cutoff
regularization we show that for $d \ge 4$ the renormalized scattering
amplitude is trivial. In contrast, dimensional regularization can
yield a nontrivial scattering amplitude for odd dimensions greater
than or equal to five. We also consider a potential consisting of a
delta function plus the derivative-squared of a delta function in
three dimensions.  We show that the renormalized scattering amplitudes
obtained using the two regularization schemes are different.  Moreover
we find that in the cutoff-regulated calculation the effective range
is necessarily negative in the limit that the cutoff is taken to
infinity. In contrast, in dimensional regularization the effective
range is unconstrained. We discuss how these discrepancies arise from the
dimensional regularization prescription that all power-law divergences
vanish. We argue that these results demonstrate that dimensional
regularization can fail in a non-perturbative setting.

\end{abstract}

\section{Introduction}

The technique known as dimensional regularization (DR)~\cite{tHV72} is
the method of choice for dealing with the infinities which appear in
perturbative quantum field theory. This elegant approach preserves
symmetries, e.g. gauge invariance and chiral symmetry, while
eliminating power-law divergences and isolating all logarithmic
divergences.  It is natural to ask why the neglect of power-law
divergences in DR is justified. In fact, the
Bogoliubov-Parasuik-Hepp-Zimmermann (BPHZ) renormalization scheme, in
which subtractions are applied directly to the integrand in divergent
expressions for loop graphs, allows perturbative renormalization to
proceed without the specification of {\em any} regularization
scheme~\cite{Co84}. Thus, given Wilson's proof that DR is a unique
procedure which is consistent with ordinary integration for finite
integrals~\cite{Wi73} it follows that, after renormalization,
perturbative calculations using DR yield results identical to those
obtained using other forms of regularization.  However, it has {\em
not} been shown that DR is equivalent to these other forms of
regularization in non-perturbative calculations.

In this paper, we discuss some examples involving the non-perturbative
regularization and renormalization of delta-function potentials in
non-relativistic quantum mechanics~\cite{BF85,GT91,MT94}. In some of
these examples DR gives physical results which differ from those
obtained using Pauli-Villars or cutoff regularization.  Our examples
explicitly illustrate the way in which DR can fail in a
non-perturbative context. They are also potentially of practical
significance, since delta-function potentials of the type discussed
here appear in effective field theory attempts to describe the
nucleon-nucleon interaction for momenta well below the pion
mass~\cite{We90,We91,Ka96}. Indeed, the issue of whether a series of
delta functions and their derivatives can systematically model a more
fundamental potential at low momenta is inextricably linked with
regularization, and will be discussed elsewhere~\cite{Be97}. The
potentials discussed in this paper will be treated as exact and not as
the leading terms in an expansion.

Our examples involve solving the Schr\"odinger (or Lippmann-Schwinger)
equation for the $S$-wave scattering of a spinless particle off a 
potential which is the sum of delta functions and derivatives of
delta functions. Here we consider potentials of the form
\begin{equation}
\langle x|V_d^{(n)}|x' \rangle=[C + C_2 ({\bf\nabla}^2 + {\bf\nabla}'^2) + C_4
({\bf\nabla}^4 + {\bf\nabla}'^4) + C_{22} {\bf\nabla}^2 {\bf\nabla}'^2 + 
\ldots] \delta^{(d)}(x-x') \delta^{(d)}(x),
\label{eq:V}
\end{equation}
where $x$ and $x'$ are $d$-dimensional Euclidean vectors. The
potential $V_d^{(n)}$ includes up to $n$ derivatives of the delta
function.  

\begin{figure}[h]
   \vspace{0.5cm}
   \epsfysize=2cm
   \centerline{\epsffile{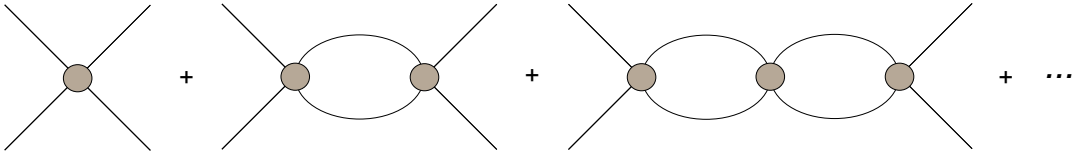}}
   \centerline{\parbox{11cm}{\caption{\label{fig1}
The Born series for $V_d^{(0)}$, which is represented
by the shaded blob. 
  }}}
\end{figure}

It is important to recognize that if the coefficients $C,C_2,\ldots$
are finite then the Hamiltonian $H$, defined by 
\begin{equation}
H=\frac{\hat{p}^2}{2 \mu} + V_d^{(n)},
\end{equation}
where $\mu$ is the mass of the particle, is not meaningful in any
integer dimension $d > 1$. One may see this by studying the Born
series for $V_d^{(0)}$. The $N$th term in this series can
be represented as an $N-1$-loop diagram. (See Fig.~\ref{fig1}.)
Solving the Lippmann-Schwinger equation involves summing this entire
series. If $d>1$ then for any finite value of the coefficient $C$ the
graphs shown in Fig.~\ref{fig1} are divergent for all $N \geq 2$.
So, in general, only after regularization and renormalization will
$V_d^{(n)}$ yield finite results.

One way to regularize and renormalize is as follows. First, all
divergent integrals are regulated by the introduction of a sharp or
smooth cutoff, thereby making the otherwise-divergent Born series
finite. This now-finite scattering amplitude is then renormalized by
choosing the coefficients to be appropriate functions of the cutoff.
These functions are chosen in such a way that certain physical
observables are reproduced.  Finally, by taking the limit of the
resulting amplitude when the cutoff goes to infinity, one recovers
results corresponding to the original potential. In other words, one
obtains finite results by choosing the coefficients $C,C_2,\ldots$ to
go to zero in a specific way.  All cutoff schemes yield identical
results for the renormalized amplitude derived from $V_d^{(n)}$,
provided that the cutoff is taken to infinity.

Such schemes represent an obvious prescription for obtaining a finite
scattering amplitude from the potential (\ref{eq:V}).  As noted above,
solving the Schr\"odinger equation is intrinsically non-perturbative,
involving the iteration of the potential to all orders. If such a
calculation were done order-by-order in the loop expansion then, since
this is a non-renormalizable theory, an infinite number of
counterterms would have to be introduced into the amplitude. On the
other hand, results may be obtained upon the introduction of a finite
number of counterterms in the potential. However, if such
non-perturbative renormalization is used the equivalence of different
regularization schemes cannot be taken for granted. Thus, a natural
question is whether using DR to render the integrals in the Born
series finite would, once renormalization is performed, lead to the
same scattering amplitude.
		
In Sec.~\ref{sec-dddf} we discuss this question for $V_d^{(0)}$, a
$d$-dimensional delta function.  A wide variety of regularization
schemes have been used to give this potential
meaning~\cite{BF85,GT91,MT94}. Moreover, Friedman has shown that for
any $d>1$ repulsive delta-function potentials are trivial, while for
any $d>3$ all delta-function potentials are trivial, i.e. the
scattering amplitude goes to zero when the regulator is removed from
the problem~\cite{Fr76,Co96}.  To our knowledge, results obtained when
DR is used to regulate the delta-function potential do not exist in
the literature for any $d \neq 3$.  We find that the renormalized
amplitudes obtained using the DR and cutoff schemes disagree in any
odd dimension $d \geq 5$. There the renormalized scattering amplitudes
obtained from the two methods of regularization are {\em
inequivalent}. After renormalization, the calculation using a cutoff
predicts a trivial scattering amplitude (in accord with Friedman's
theorem) while the calculation using DR predicts a constant amplitude
which in general is not constrained to vanish.

Of course, the five-dimensional delta function, while clearly
illustrating possible disagreements between DR and cutoff schemes, is
not otherwise intrinsically interesting. A problem of more relevance
to physical calculations is studied in Sec.~\ref{sec-df2}, where we
consider potentials containing two derivatives of the delta function
in $d=3$, i.e. $V^{(2)}_3$ of Eq.~(\ref{eq:V}). This potential has
been used in effective field theory treatments of the nucleon-nucleon
interaction~\cite{Ka96}.  In this problem DR and cutoff schemes
predict different physical amplitudes.  The on-shell $T$ matrix
obtained using DR is
\begin{equation}
T^{\rm on}_{\rm DR}(E)=-\frac{2 \pi}{\mu}\frac{1}{1/(-a - r_e a^2 \mu E) 
- i \sqrt{2 \mu E}},
\end{equation}
while that found using cutoff regularization is, in the infinite
cutoff limit,
\begin{equation}
T^{\rm on}_{\rm cutoff}(E)=-\frac{2 \pi}{\mu} \frac{1}
{-1/a + r_e \mu E - i\sqrt{2 \mu E}}; \,\,\,r_e\leq 0,
\end{equation}
where $a$ and $r_e$ are the scattering length and effective range.
Observe that the scattering amplitudes obtained using the two schemes
have different functional forms. Moreover, the two results do not
apply under the same conditions. In DR it appears that the amplitude
can always be renormalized regardless of the magnitude or sign of
$r_e$ and $a$. In contrast, in the cutoff regularization scheme the
non-linear renormalization conditions can only be satisfied if the
effective range is negative.  In Sec.~\ref{sec-wb} we discuss a
generalization of these results which is based on a result derived by
Wigner~\cite{Wi55,CP96}.  We show that, no matter what potential
$V_3^{(n)}$ is used, in the cutoff scheme the renormalized on-shell
amplitude always obeys
\begin{equation}
\frac{\partial}{\partial E}
\mbox{Re}\left(\frac{1}{T^{\rm on}(E)}\right)\leq 0.
\end{equation}
No such constraint applies to the on-shell amplitude generated by
DR.

The source of the inequivalence between these results obtained using
different regularization methods is easily diagnosed. It is DR's
prescription that all power-law divergences be discarded. By contrast,
in cutoff regularization these divergences are retained and must be
dealt with by renormalization.  In perturbative calculations this
difference in the treatment of power-law divergences is irrelevant
after renormalization, since to a given order in perturbation theory
such divergences may always be absorbed into counterterms~\footnote{
Sometimes symmetries require that the sum of contributions from power
law divergences and their counterterms vanish.  DR is particularly
convenient for such cases.  Consequently, it is well suited to both
chiral theories and gauge theories.}. The key point is that DR often
involves more than {\em just} regularizing integrals. It also does a
certain amount of renormalization by automatically subtracting all
power-law divergences. There is therefore an implicit assumption that
these effects can always be absorbed into counterterms. Our examples
show that in certain non-perturbative calculations this partial
renormalization that is automatically and implicitly done in DR can
lead to unphysical results. This is because in non-perturbative
problems, the renormalization conditions relating bare parameters to
experimental data are generally non-linear, and it is not always
possible to absorb all power-law divergences into counter terms. These
issues are discussed and our conclusions drawn in
Sec.~\ref{sec-drdiff}.

\section{The delta-function potential in $d$ dimensions}

\label{sec-dddf}

In this section we discuss scattering from the delta-function
potential in $d$ dimensions.  We solve the Lippmann-Schwinger equation
for this potential in momentum space at energies $E \geq 0$, using
both a sharp cutoff and DR, and demonstrate that the resultant
renormalized amplitudes are different in any odd dimension $d>3$. The
generalization of our arguments to a smooth momentum-space cutoff is
straightforward. Solutions for the delta-function potential where the
regularization is done using position-space cutoffs may be found in
Refs.~\cite{GT91,MT94}.  Discussions of the rigorous definition of
such potentials via the construction of self-adjoint extensions of the
Laplacian on the space of smooth functions compactly supported away
from the origin appear in Refs.~\cite{Ja91,Al88,Fe95}. Here we avoid
these mathematical questions and simply solve the Lippmann-Schwinger
equation for these potentials using the regularization and
renormalization procedure defined in the Introduction.

We wish to solve the Lippmann-Schwinger equation:
\begin{equation}
T(p',p;E)=V(p',p) + \int \frac{d^dk}{(2 \pi)^d} \, V(p',k) 
\frac{1}{E^+ - \frac{k^2}{2 \mu}} T(k,p;E),
\label{eq:LSE}
\end{equation}
where $E^+\equiv E+i\epsilon$, with $E \geq 0$, and $\mu$ is the
reduced mass in the two-body system. In momentum space, the
delta-function potential is given by
\begin{equation}
V(p',p)=C.
\end{equation}
{}From Eq.~(\ref{eq:LSE}) it is apparent that the corresponding
T-matrix has no momentum dependence, i.e.,
\begin{equation}
T(p',p;E)=T^{\rm on}(E).
\end{equation}
It follows that
\begin{equation}
\frac{1}{T^{\rm on}(E)}=\frac{1}{C} - I(E),
\end{equation}
where

\begin{eqnarray}
I(E) &\equiv& \int \frac{d^dk}{(2 \pi)^{d}} \, \frac{1}{E^+ -
\frac{k^2}{2 \mu}},\label{eq:IEdef}\\ &=& \frac{4 \pi^{d/2}
\mu}{\Gamma(d/2) (2 \pi)^d} \int_0^\infty dk \frac{k^{d-1}} {2 \mu E^+
- k^2}.
\end{eqnarray}
We have done the angular integration, thereby obtaining a factor equal
to the surface area of the unit sphere in $d$ dimensions. Using the
identity

\begin{equation}
\int_0^\infty dk \, f(k) \frac{1}{2 \mu E^+ - k^2} = P \int_0^\infty dk \, 
f(k) \frac{1}{2 \mu E - k^2} - i \frac{\pi}{2 \sqrt{2 \mu E}} f(\sqrt{2 \mu E}),
\end{equation}
where $P$ denotes a principal value integral, we see that
\begin{equation}
I(E)=\frac{4 \pi^{d/2} \mu}{\Gamma(d/2) (2 \pi)^d} P \int_0^\infty dk \,
\frac{k^{d-1}} {2 \mu E - k^2} - i \frac{2 \pi^{d/2+1} \mu}
{\Gamma(d/2) (2 \pi)^d} (2 \mu E)^{d/2 - 1}.
\label{eq:IE}
\end{equation}
The real part of the integral $I(E)$ is clearly divergent for all $d
\geq 2$, and so requires some form of regularization.  Suppose that we
calculate this integral by isolating all divergent pieces and
introducing a momentum cutoff $\beta$.  In order to do this we make
use of the relation:
\begin{equation}
P\int_0^\infty dk \, \frac{k^{n}}{2 \mu E - k^2}=-\int_0^\infty dk \,
k^{n-2} + 2 \mu E \, P \int_0^\infty dk \, \frac{k^{n-2}}{2 \mu E -
k^2},
\label{eq:reln}
\end{equation}
which holds for all $n \geq 2$. This leads to the following series expressions:
\begin{equation}
P\int_0^\infty dk \, \frac{k^{d-1}}{2 \mu E - k^2}=\left\{\begin{array}{ll}
-\frac{\beta^{d-2}}{d-2} - 2 \mu E \frac{\beta^{d-4}}{d-4} - \ldots
- (2 \mu E)^{(d-3)/2} \beta & \mbox{$d$ odd}\\
-\frac{\beta^{d-2}}{d-2} - \ldots
- (2 \mu E)^{d/2-2} \frac{\beta^2}{2} + \frac{1}{2} (2 \mu E)^{d/2-1} \,
\log(\frac{2 \mu E}{\beta^2 - 2 \mu E}) & \mbox{$d$ even,} 
\end{array} \right.
\label{eq:COP}
\end{equation}
provided that $\beta^2 > 2 \mu E$.
Note that in the case where $d$ is odd we have not regulated the integral
\begin{equation}
P\int_0^\infty dk \, \frac{1}{2 \mu E - k^2}, 
\end{equation}
since this convergent integral is equal to zero. If a smooth cutoff is
used these results still hold, but different numerical coefficients
reside in front of the terms in the series in Eq.~(\ref{eq:COP}). The
argument which follows is completely unaffected by this change.

By contrast, the standard methods of evaluation used in dimensional 
regularization (see, e.g.~\cite{IZ84}) lead to:
\begin{equation}
\mbox{Re}[ I(E) ]=
-\frac{2 \mu}{(4 \pi)^{d/2}}\, \mbox{Re}[(-2 \mu E)^{d/2-1}] \, \Gamma(1-d/2). 
\label{eq:DRP}
\end{equation}
Note that for any $E>0$ and any odd $d$ this yields
\begin{equation}
\mbox{Re}[ I(E) ]=0 .
\end{equation}

In order to have either of the results (\ref{eq:COP}) or
(\ref{eq:DRP}) for $I(E)$ yield physical quantities we must
renormalize the amplitude $T(p',p;E)$. Since for this potential
$T(p',p;E)=T^{\rm on}(E)$ we choose some energy $E_0 \geq 0$ and write
the renormalization condition as
\begin{equation}
T(p',p;E_0)=t,
\label{eq:renromcond}
\end{equation}
where $t$ is finite.

First, consider the case $d=2$. In this case cutoff regularization
gives:
\begin{equation}
\mbox{Re}\left(\frac{1}{T^{\rm on}(E)}\right)=
\frac{1}{C_{\rm cutoff}} + \frac{\mu}{2 \pi}[\log(\beta^2 - 2 \mu E) 
- \log(2 \mu E)].
\end{equation}
So choose 
\begin{equation}
\frac{1}{C_{\rm cutoff}}=\mbox{Re}\left(\frac{1}{t}\right) - 
\frac{\mu}{2 \pi}[\log(\beta^2 - 2 \mu E_0) - \log(2 \mu E_0)].
\end{equation}
Then, in the limit as $\beta \rightarrow \infty$:
\begin{equation}
\frac{1}{T^{\rm on}(E)}=\mbox{Re}(\frac{1}{t}) + \frac{\mu}{2 \pi}\log(\frac{E_0}{E})
+ i \frac{\mu}{2}.
\label{eq:renorm2damp}
\end{equation}

This is to be compared to what occurs in DR. There
\begin{equation}
\mbox{Re} \left(\frac{1}{T^{\rm on}} \right)=\frac{1}{C_{\rm DR}} + 
\frac{\mu}{2 \pi} \lim_{d \rightarrow 2} \Gamma(1 - d/2) \mbox{Re}[(-2 \mu E)^{d/2 - 1}].
\end{equation}
Thus,
\begin{equation}
\mbox{Re} \left(\frac{1}{T^{\rm on}}\right)=\frac{1}{C_{\rm DR}} +
\frac{\mu}{2 \pi} \lim_{d \rightarrow 2} \left[\left(\frac{1}{1-d/2} -
\gamma + \ldots\right)\left( 1 + (d/2-1)\log(2 \mu E)\right)\right],
\end{equation}
where $\gamma$ is the Euler number. So, we choose:
\begin{equation}
\frac{1}{C_{\rm DR}}=\mbox{Re} \left(\frac{1}{t}\right) -
\frac{\mu}{2 \pi} \left[ \lim_{d \rightarrow 2} \frac{1}{1-d/2}
- \log(2 \mu E_0) -  \gamma \right],
\end{equation}
and recover Eq.~(\ref{eq:renorm2damp}).

Next, take the case $d=3$. There cutoff regularization gives
\begin{equation}
\mbox{Re} \left(\frac{1}{T^{\rm on}(E)}\right)=\frac{1}{C_{\rm cutoff}} + 
\frac{\mu \beta}{\pi^2}, 
\end{equation}
while DR yields
\begin{equation}
\mbox{Re} \left(\frac{1}{T^{\rm on}(E)}\right)=\frac{1}{C_{\rm DR}}.
\end{equation}
Thus, by choosing $C$ appropriately, in both cases the result
\begin{equation}
\frac{1}{T^{\rm on}(E)}=\mbox{Re} \left(\frac{1}{t}\right) 
+ i \frac{\mu \sqrt{2 \mu E}}{2 \pi},
\end{equation}
derived by Weinberg~\cite{We91} and Kaplan {\it et al.}~\cite{Ka96} is
obtained. Therefore, in $d=2$ and $d=3$, both regularization schemes
lead to equivalent physical results. 

Note that in $d=2$ and $d=3$ ${C_{\rm cutoff}}$ must be negative as
the cutoff is taken to infinity.  This corresponds to an attractive
bare potential. Therefore, if the bare potential is repulsive, the
scattering amplitude must be trivial, in accord with Friedman's
theorem~\cite{Fr76,Co96}. In $d=2$, DR is consistent with Friedman's
theorem if the coefficient of the $1/(d-2)$ pole is positive as
$d\rightarrow 2$.  In $d=3$, DR places no constraint on the bare
potential. However, the seriousness of this inconsistency with
Friedman's theorem is open to question, since the sign of the bare
potential is not necessarily physically significant.

Next consider the calculation of $I(E)$ in any even dimension $d \geq
4$. If cutoff regularization is used then Eq.~(\ref{eq:COP}) indicates
that the resulting $I(E)$ contains an energy-dependent divergence.
Consequently, although the resulting amplitude may be made finite at
energy $E_0$ by choosing $1/C_{\rm cutoff}$ appropriately, it is
impossible to make $1/T^{\rm on}$ finite for any other energy.
Therefore $1/T^{\rm on}$ diverges unless we require that $C_{\rm
cutoff}$ vanishes, and we recover the result expected from Friedman's
theorem; the scattering amplitude is trivial~\cite{Fr76}.

In DR in even dimensions, say $d=2 w$,

\begin{equation}
\frac{1}{T^{\rm on}_{\rm DR}(E)}=\frac{1}{C_{\rm DR}} + \frac{2 \mu}
{(4 \pi)^{w}} (-2 \mu E)^{w-1} \Gamma(1-w) + i \frac{2 \pi \mu}
{\Gamma(w) (4 \pi)^{w}} (2 \mu E)^{w - 1}.
\end{equation}
Since $\Gamma(1-w)$ is divergent for all positive integers 
for all $w \geq 2$ an energy-dependent divergence appears here too.
Therefore for even dimensions $d \geq 4$ both regularization
schemes predict a trivial scattering amplitude:
\begin{equation}
T^{\rm on}(E)=0.
\end{equation}

The case of odd dimensions $d \geq 5$ is more interesting. Once again, in
cutoff regularization the integral $I(E)$ contains an energy-dependent
divergence and so the renormalized amplitude is trivial, 	
in accord with Friedman's theorem:
\begin{equation}
T^{\rm on}_{\rm cutoff}(E)=0.
\label{eq:cutoff}
\end{equation}
By contrast, in DR the real part of the integral $I(E)$ is zero in
each odd dimension $d \geq 3$. So,

\begin{equation}
\frac{1}{T^{\rm on}_{\rm DR}(E)}={\frac{1}{C_{\rm DR}} 
+ i \frac{2 \pi \mu} {\Gamma(d/2) (4 \pi)^{d/2}} (2 \mu E)^{d/2 - 1}}.
\label{eq:DR}
\end{equation}
Renormalization can now be performed, leading to a nontrivial
scattering, by choosing $1/C_{\rm DR}=\mbox{Re}(1/t)$. The cutoff and
dimensionally regularized theories are therefore inequivalent. Indeed,
the amplitude found using DR corresponds to one obtained from a
Schr\"odinger equation with a non-Hermitian pseudo-potential, as shown
explicitly in five dimensions by Grossman and Wu~\cite{GW84}.

However, the sense in which the two calculations starting with the
potential $V^{(0)}$ are inequivalent is somewhat limited.  After all,
one could choose $C_{\rm DR}=0$ in the dimensionally regularized
theory and so obtain a trivial amplitude there as well. Nonetheless,
there is an important way in which the two approaches are
inequivalent---in one case one concludes that the theory is capable of
yielding a finite, nonzero scattering amplitude and in the other case
one concludes that it is not.  In the next section we present a more
compelling example of inequivalence, where the two regularization
schemes give different, finite, nonzero scattering amplitudes.

\section{A delta function and its derivative-squared in three dimensions}

\label{sec-df2}

In this section we consider the example of the potential $V_3^{(2)}$,
i.e.~a potential in three dimensions which consists of a delta
function plus two derivatives thereof. We show that for this potential
cutoff regularization and DR give finite inequivalent results for the
renormalized $S$-wave scattering amplitude.

In momentum space this potential may be written
\begin{equation}
V(p',p)=C + C_2 (p^2 + p'^2).
\end{equation}
We want to insert this into the Lippmann-Schwinger equation
(\ref{eq:LSE}) and solve for $d=3$. One way to do this is
to observe that $V$ may be written as a two-term separable potential
\begin{equation}
V(p',p)=\sum_{i,j=0}^1 p'^{2i} \lambda_{ij} p^{2j},
\end{equation}
where the matrix ${\bf \lambda}$ is
\begin{equation}
\{\lambda_{ij}\}_{i,j=0}^1=\left( \begin{array}{cc} C & C_2\\
						    C_2 &0
				  \end{array} \right).
\end{equation}
The solution to the Lippmann-Schwinger equation then takes the form
\begin{equation}
T(p',p;E)=\sum_{i,j=0}^1 p'^{2i} \tau_{ij}(E) p^{2j},
\label{eq:Tmat}
\end{equation}
where ${\bf \tau}$ obeys the matrix equation:
\begin{equation}
{\bf \tau}(E)={\bf \lambda} + {\bf \lambda} {\bf {\cal I}}(E) {\bf \tau}(E),
\label{eq:tau}
\end{equation}
with
\begin{equation}
{\bf {\cal I}}(E)=\left( \begin{array}{cc} 
\int \frac{d^3k}{(2 \pi)^3} \frac{1}{E^+ - \frac{k^2}{2 \mu}} & 
\int \frac{d^3k}{(2 \pi)^3} \frac{k^2}{E^+ - \frac{k^2}{2 \mu}}\\
\int \frac{d^3k}{(2 \pi)^3} \frac{k^2}{E^+ - \frac{k^2}{2 \mu}} & 
\int \frac{d^3k}{(2 \pi)^3} \frac{k^4}{E^+ - \frac{k^2}{2 \mu}}
\end{array} \right).
\end{equation}
Now using the relation (\ref{eq:reln}) we see that:
\begin{equation}
{\bf {\cal I}}(E)=\left( \begin{array}{cc}
I(E) & I_3 + 2 \mu E \, I(E)\\
I_3 + 2 \mu E \, I(E) & I_5 + 2 \mu E \, I_3 + 4 \mu^2 E^2 \, I(E),
\end{array} \right),
\end{equation}
with Eq.~(\ref{eq:IE}) for $d=3$ defining $I(E)$ and
\begin{equation}
I_5 \equiv -2 \mu \int \frac{d^3k}{(2 \pi)^3} k^2; \quad
I_3 \equiv -2 \mu \int \frac{d^3k}{(2 \pi)^3}.
\end{equation}

We are now in a position to solve the algebraic equation
(\ref{eq:tau}) for ${\bf \tau}(E)$. This gives
\begin{equation}
{\bf \tau}(E)=\frac{1}{\det {\bf \tau}^{-1}} \left( \begin{array}{cc}
-\frac{C}{C_2^2} - I_5 - 2 \mu E I_3 - 4 \mu^2 E^2 I(E) & I_3 + 
2 \mu E \, I(E) -\frac{1}{C_2}\\ 
I_3 + 2 \mu E \, I(E) - \frac{1}{C_2} & -I(E)
\end{array} \right),
\end{equation}
with,
\begin{equation}
\det {\bf \tau}^{-1}=\left[\frac{C}{C_2^2} I(E) + I_5 I(E) - 
2 \mu E \, I_3 I(E) - (\frac{1}{C_2} - I_3)^2 + 4 \mu E I(E)
\frac{1}{C_2} \right]
\end{equation}
When this expression is inserted into Eq.~(\ref{eq:Tmat}), we find for the 
on-shell t-matrix:

\begin{equation}
\frac{1}{T^{\rm on}(E)}=\frac{(C_2 I_3 -1)^2}{C + C_2^2 I_5 + 2 \mu E C_2 (2 -
C_2 I_3)} - I(E).
\label{eq:Tonexp}
\end{equation}
The integrals $I_3$, $I_5$ and $\mbox{Re}(I(E))$ are all divergent, and so
this amplitude requires regularization and renormalization. Indeed,
{\it a priori} it is not clear that renormalization will be possible,
since we have three distinct infinities and only two counterterms.

In fact this renormalization can be carried a certain distance without
making reference to any particular regularization scheme. We choose as
renormalization parameters the experimental values of the scattering
length, $a$, and the effective range, $r_e$. In other words, we fix $C$
and $C_2$ by demanding that
\begin{equation}
\frac{1}{T^{\rm on}(E)}=-\frac{\mu}{2 \pi}\left(-\frac{1}{a} +
r_e \mu E + O( \mu^2 E^2) - i \sqrt{2 \mu E} \right).
\label{eq:reexp}
\end{equation}
Comparing this to Eq.~(\ref{eq:Tonexp}) shows that the 
imaginary parts automatically agree (as is guaranteed by the unitarity
of the Lippmann-Schwinger equation). Equating the real parts at $E=0$ yields
\begin{equation}
\frac{ \mu}{2 \pi a}=\frac{(C_2 I_3 -1)^2}{C + C_2^2 I_5} - I(0).
\end{equation}
Noting that,

\begin{equation}
\mbox{Re}(I(E))=I(0)\equiv I_1=-2 \mu \int \frac{d^3k}{(2 \pi)^3} 
\frac{1}{k^2},
\end{equation}
this gives

\begin{equation}
\mbox{Re}\left(\frac{1}{T^{\rm on}(E)}\right)=
\frac{\mu/(2 \pi a) - 2 \mu E \, I_1 A}{1 + 2 \mu E \, A},
\label{eq:Tsimple}
\end{equation}
with

\begin{equation}
A \equiv (\frac{\mu}{2 \pi a} + I_1) \frac{C_2 (2 - C_2 I_3)}{(C_2
I_3 - 1)^2}.
\label{eq:Adef}
\end{equation}
From Eqs.~(\ref{eq:reexp}), (\ref{eq:Tsimple}), and (\ref{eq:Adef}) the 
renormalization condition relating $C_2$ to the physical parameter $r_e$ 
may be economically expressed as a condition on $A$:

\begin{equation}
A=\frac{\mu r_e}{4 \pi}\left(I_1 + \frac{\mu}{2 \pi a}\right)^{-1}.
\label{eq:Acondn}
\end{equation}
Let us consider the structure of this renormalization condition and
the resulting amplitude (\ref{eq:Tsimple}) for both DR and cutoff
regularization.

First, we regularize using DR. There   

\begin{equation}
I_1=I_3=I_5=0.
\end{equation} 
Consequently, the renormalization condition (\ref{eq:Acondn}) becomes

\begin{equation}
A=\frac{1}{2} a r_e,
\end{equation}
with $A=C_2 \mu/\pi a$. The term in the numerator of
Eq.~(\ref{eq:Tsimple}) disappears, leading to

\begin{equation}
\frac{1}{T_{\rm DR}^{\rm on}(E)}=-\frac{\mu}{2 \pi}
\left(\frac{1}{-a - a^2 r_e \mu E} - i\sqrt{2 \mu E} \right)
\label{eq:TonDR}
\end{equation} 
for all $a$ and $r_e$, as found by Kaplan {\it et al.}~\cite{Ka96}. 

By contrast, if a cutoff is used when the renormalization
condition (\ref{eq:Acondn}) is rewritten in terms of $C_2$ we have

\begin{equation}
\frac{\mu r_e}{4 \pi}=\left(\frac{\mu}{2 \pi a} + I_1\right)^2
\left[\frac{1}{(C_2 I_3 - 1)^2 I_3} - \frac{1}{I_3}\right].
\label{eq:recondn}
\end{equation}
In any scheme where a cutoff $\beta$ is used to regulate the theory we
have, on dimensional grounds:
\begin{equation}
\frac{I_1^2}{I_3} \sim \frac{1}{\beta},
\label{eq:scaling}
\end{equation}
and thus the second term in Eq.~(\ref{eq:recondn}) disappears in the
limit that the cutoff is taken to infinity. Consequently, as $\beta
\rightarrow \infty$
\begin{equation}
\frac{\mu r_e}{4 \pi} \rightarrow \frac{1}{I_3} \left(\frac{I_1}{C_2 I_3
- 1}\right)^2,
\end{equation}
from which we see that $r_e \leq 0$, {\it independent of the value of
$C_2$}, provided only that $C_2$ is real, i.e. the original bare
Hamiltonian is Hermitian. It could be argued that the bare
Hamiltonian, being unobservable, is not required to be Hermitian. We
will return to the issue of the physical significance of the bare
Hamiltonian in Section~\ref{sec-drdiff} and in Ref.~\cite{Be97}. 

The important point is that, from Eqs.~(\ref{eq:recondn}), (\ref{eq:Adef}) and
(\ref{eq:Tsimple}) we see that the cutoff scheme yields an amplitude
which, as the cutoff is taken to infinity, becomes

\begin{equation}
\frac{1}{T_{\rm cutoff}^{\rm on}(E)}=-\frac{\mu}{2 \pi}\left(-\frac{1}{a} + 
r_e \mu E - i \sqrt{2 \mu E} \right);\qquad  r_e \leq 0.
\label{eq:cutoff2}
\end{equation}
Note that the type of cutoff regularization used here is irrelevant.
In any scheme where the cutoff carries powers of momentum the behavior
(\ref{eq:scaling}) occurs as the cutoff is removed, and we are led to
the form (\ref{eq:cutoff2}) for the amplitude. This has a functionally
different form to that obtained in Eq.~(\ref{eq:TonDR}) using DR. This
means that even if $r_e \leq 0$ and the renormalization condition
(\ref{eq:Acondn}) can be satisfied, the DR results are still different
from those based on a cutoff.

It is clear why DR leads to a different renormalized amplitude.  The
cause is the DR prescription of discarding all power-law
divergences. The different results obtained for the form of the
amplitude may be thought of as arising from the way the ratio $A$
behaves in the two different regularization schemes.  In DR $I_1
\equiv 0$, and $A=\frac{1}{2} a r_e$, so the term in the
numerator of Eq.~(\ref{eq:Tsimple}) disappears, thus leading to
(\ref{eq:TonDR}).  However, in a cutoff scheme, as $\beta \rightarrow
\infty$, $I_1 A \rightarrow \frac{\mu}{4 \pi} r_e$, and so the term
in the denominator of Eq.~(\ref{eq:Tsimple}) disappears, leading to
the form (\ref{eq:cutoff2}). Moreover, in the cutoff case as $\beta
\rightarrow \infty$ $A$ has a fixed sign, and so if $r_e > 0$
renormalization is simply impossible. Since the ratio $I_1^2/I_3$ has
no meaning in DR the sign of $A$, and hence that of $r_e$, is
unconstrained in that approach.

\section{The Wigner bound on the amplitude generated by $V_3^{(2N)}$}

\label{sec-wb}

We have seen that if cutoff regularization is used to give meaning to
$V_3^{(2)}$ renormalization can only be performed if the effective
range is negative. One might think that this constraint is a special
feature of the calculation with $V_3^{(2)}$, but here we show that it
is a result which holds no matter how many derivatives of the delta
function are included in the potential (\ref{eq:V}). We do this by
making a connection with the bound on $r_e$ originally derived by
Wigner~\cite{Wi55}, rederived by Fewster~\cite{Fe95}, and discussed in
Ref.~\cite{CP96}. There it was shown that, for any energy-independent
potential $V(r',r)$ which obeys
\begin{equation}
V(r',r)=0 \mbox{ for all $r,r' > R$},
\end{equation}
the effective range is bounded by
\begin{equation}
r_e \leq 2 \left(R - \frac{R^2}{a} + \frac{R^3}{3 a^2}\right).
\label{eq:WB}
\end{equation}
It was also shown that this bound applies even if the potential does
not go strictly to zero, but merely decreases fast enough for the wave
function to approach the asymptotic solution sufficiently quickly.

In order to make the bound of Ref.~\cite{CP96} apply to the problem we
have been discussing here we need to show two things. First, we must
show that the position-space arguments of Ref.~\cite{CP96} can be
translated into momentum space.  Second, we must prove that, in the
limit that the regulator is removed (i.e.~the regulator mass goes to
infinity), regulating the potential and then subsequently iterating it
via the Lippmann-Schwinger equation is equivalent to formally
iterating the interaction and then regulating all divergent integrals
using a cutoff.

To explain the first point we consider the class of potentials
\begin{equation}
V_\beta(p',p)=g(\frac{p'^2}{\beta^2}) \left(\sum_{i,j=0}^{N} {p'}^{2i} 
\lambda_{ij} p^{2j}\right) g(\frac{p^2}{\beta^2}).
\label{eq:regpot}
\end{equation}
Here $\beta$ is the cutoff parameter of the (sharp or smooth) cutoff
function $g$. This function $g(x^2)$ obeys $g(0)=1$ and $g(x^2)
\rightarrow 0$ faster than $\frac{1}{x^{N + 1/2}}$ as $x \rightarrow
\infty$. (For instance, $g(x^2)=\exp(-x^2)$ is an acceptable choice.)
In the limit $\beta \rightarrow \infty$ this class of potentials
formally approaches the class of potentials we are considering in
Eq.~(\ref{eq:V}) with $M_N$ of the $\lambda_{ij}$'s nonzero where

\begin{equation}
M_N=\left\{ \begin{array}{ll}
\frac{N^2}{4} + N + \frac{3}{4} & \mbox{ if $N$ is odd}\\
\frac{N^2}{4} + N + 1           & \mbox{ if $N$ is even.}
\end{array} \right.
\label{eq:MN}
\end{equation}
Therefore once an amplitude is calculated using $V_\beta$ $M_N$
renormalization conditions are needed.  Renormalization can be
achieved by fixing $\beta$ and then calculating the $\lambda_{ij}$s
needed to reproduce the effective range expansion to order $M_N$ for
that value of $\beta$.  Below we show that as $\beta \rightarrow
\infty$ the amplitude obtained by this new procedure is identical to
that derived via the method used in the previous section.

But first we show that the Wigner bound on the effective range (\ref{eq:WB})
applies to the potential $V_\beta(p',p)$. Equation (\ref{eq:WB}) was derived
in position space. Taking the Fourier transform of (\ref{eq:regpot}) we
get

\begin{equation}
V_\beta(r',r)=\tilde{g}(r'^2 \beta^2) \left[\sum_{i,j=0}^{N}
(-\stackrel{\leftarrow}{\nabla'}{}^2)^i \lambda_{ij}
(-\stackrel{\rightarrow}{\nabla}{}^2)^j\right] {\tilde g}(r^2
\beta^2),
\end{equation}
where $\tilde{g}$ is the Fourier transform of $g(k^2)$ with respect to
$\vec{k}$. For any $g$ which, for arbitrary $N$, obeys the conditions
given above no finite $R$ exists for which $\tilde{g}$ goes strictly
to zero for all $r,r'>R$. However, by choosing $g$ appropriately we
can, for $\beta$ large enough, make $V_\beta(r,r')$ arbitrarily close
to zero.  Hence the wave function can be made to approach the
asymptotic solution rapidly enough for the Wigner bound to apply, with
the range $R \rightarrow 0$ as $\beta \rightarrow \infty$.  Thus, as
$\beta \rightarrow \infty$ Eq.~(\ref{eq:WB}) yields:
\begin{equation}
r_e \leq 0,
\end{equation}
for the renormalized amplitude generated by $V_\beta$ {\it regardless
of the value of $N$}.

So, all that is left for us to prove is that iterating and then using
cutoff regularization is equivalent to regulating via, e.g.
Eq.~(\ref{eq:regpot}), and then iterating.  That is, we must show
that, once the cutoff is taken to infinity, iterating the regularized
potential $V_\beta$ and renormalizing the coefficients is equivalent
to formally iterating the potential $V_3^{(2N)}$, regulating all
divergent integrals using a cutoff, and then renormalizing.  To do
this let us consider an unregulated potential $V_3^{(2N)}$:
\begin{equation}
V(p',p)=\sum_{i,j=0}^{N} {p'}^{2i} \lambda_{ij} p^{2j}.
\label{eq:VN}
\end{equation}
The t-matrix generated by this interaction has the form
\begin{equation}
T(p',p;E)=\sum_{i,j=0}^{N} {p'}^{2i} \tau_{ij}(E) p^{2j},
\end{equation}
where ${\bf \tau}(E)$ is to be found using Eq.~(\ref{eq:tau}) with the matrix
$\{{\cal I}\}_{i,j=0}^N$ defined by
\begin{equation}
{\cal I}_{ij}(E)=\int \frac{d^3k}{(2 \pi)^3} \frac{k^{2(i+j)}}{E^+ 
- \frac{k^2}{2 \mu}}.
\end{equation}
Now using the recursion relation (\ref{eq:reln}) we observe that
\begin{equation}
{\cal I}_{ij}(E)=\tilde{\cal I}_{ij}(E) + (2 \mu E)^{i+j} I(E).
\end{equation}
These arguments are similar to those used in solving the Lippmann-Schwinger
equation for the potential $V_\beta$. However, one main difference is that 
there:
\begin{equation}
{\cal I}_{ij}(E)=\int \frac{d^3k}{(2 \pi)^3} \frac{k^{2(i+j)}}{E^+ 
- \frac{k^2}{2 \mu}} \left[g\left(\frac{k^2}{\beta^2}\right)\right]^2.
\end{equation}
Nevertheless, since the integrals $\tilde{\cal I}_{ij}$ are all real
and divergent, they are unaffected by the order in which
regularization and iteration are performed.

Define
\begin{equation}
\tilde{\bf \tau}^{-1}(E)={\bf \lambda}^{-1} - \tilde{\bf {\cal I}}(E).
\label{eq:tildetau}
\end{equation}
It follows that
\begin{equation}
{\bf \tau}^{-1}(E)=\tilde{\bf \tau}^{-1}(E) - {\bf g}(E) I(E),
\label{eq:tildeeq}
\end{equation}
where the matrix ${\bf g}(E)$ is defined by:
\begin{equation}
\{g\}_{i,j=0}^N=(2 \mu E)^{i+j}.
\end{equation}
If we define $\tilde{T}^{\rm on}$ via
\begin{equation}
\tilde{T}^{\rm on}(E)=\sum_{i,j=0}^N (2 \mu E)^{i+j} \tilde{\tau}_{ij}(E)
\end{equation}
then from Eq.~(\ref{eq:tildeeq}) we obtain:
\begin{equation}
\frac{1}{T^{\rm on}(E)}=\frac{1}{\tilde{T}^{\rm on}(E)} - I(E).
\end{equation}
Note that since the imaginary part of $I(E)$ is given by the formula
(\ref{eq:IE}) with $d=3$, and the amplitude $\tilde{T}$ is real, this
argument shows that the amplitude $T^{\rm on}$ will always be unitary.

Equation (\ref{eq:tildetau}) implies that the amplitude
$\tilde{T}^{\rm on}$ is unaffected by the order in which
regularization and iteration are performed. Therefore, the only way
the order of these two actions can affect $T^{\rm on}(E)$ is via the
real part of the integral $I(E)$. We saw above that if regularization
is performed after iteration $\mbox{Re}(I(E))=I_1$. By contrast, if the
potential $V_\beta$ is used, we have, for $2 \mu E \ll \beta^2$,
\begin{equation}
\mbox{Re}(I(E))=I_1 - 2 \mu \beta 
\sum_{i=1}^\infty c_i \left(\frac{2 \mu E}{\beta^2}\right)^{i},
\label{eq:regI}
\end{equation}
where the numerical coefficients $c_i$ are always of order one, with
their exact value dependent upon the particular regulator used. Now
let us consider what happens when renormalization is performed.
Suppose that the renormalization conditions are chosen so as to
enforce agreement with the effective range expansion for
$\mbox{Re}(\frac{1}{T^{\rm on}(E)})$ up to order $M_N$.  Since
$\mbox{Re}(I(E))$ only appears in $\mbox{Re}\left(\frac{1}{T^{\rm
on}(E)}\right)$, $M_N$ of the terms in the sum of Eq.~(\ref{eq:regI})
can be absorbed by adjusting the unrenormalized coefficients in
$V_\beta$. The terms which are not absorbed via this step then appear
in the renormalized amplitude at any finite $\beta$.  However, in the
limit when $\beta \rightarrow
\infty$ they make no contribution to the final result.  Indeed, once
$\beta \rightarrow \infty$ none of the terms in (\ref{eq:regI}) except
$I_1$ play any role, either during or after renormalization.  Thus the
order of iteration and regularization is immaterial once the cutoff is
taken to infinity.

Consequently the Wigner bound applies to $V_\beta$ in the limit $\beta
\rightarrow \infty$ and in that limit iterating $V_\beta$ yields the
same amplitude as iterating (\ref{eq:VN}) and then using cutoff
regularization. Hence, no matter which $V^{(2N)}$ is used, in the
limit that the cutoff is removed from the theory the renormalized
amplitude can only predict a negative effective range. In fact, in the
limit that the range of the potential goes to zero Ref.~\cite{CP96}
showed that
\begin{equation}
\frac{d}{d\bar{p}^2} (\bar{p} \cot \delta(\bar{p})) \leq 0,
\label{eq:bound}
\end{equation}
for all $\bar{p}=\sqrt{2 \mu E}$. But, everything said above for the
effective range applies to higher derivatives of the real part of the
inverse amplitude. Therefore, regardless of the potential
$V^{(2N)}$ used, and regardless of the order in which cutoff regularization
and iteration are done, the bound (\ref{eq:bound}) applies to the
resultant amplitude.

But what happens if DR is used instead of cutoff regularization? To
answer this question consider Eq.~(\ref{eq:tildetau}).  Since all the
integrals $\tilde{\cal I}_{ij}$ are power-law divergent with no
finite or logarithmically-divergent part, in DR,
\begin{equation}
\tilde{\cal I}_{ij}(E)=0 \mbox{ for all $i,j=1,2,\ldots,N$}.
\end{equation}
Therefore, 
\begin{equation}
\tilde{\bf \tau}_{DR}(E)={\bf \lambda}^{-1}.
\end{equation}
As observed in the previous section $\mbox{Re}(I(E))=0$ in DR also, thus
\begin{equation}
\frac{1}{T_{DR}^{\rm on}(E)}=\frac{1}{\tilde{T}_{DR}^{\rm on}(E)} + 
i \frac{\mu \sqrt{2 \mu E}}{2 \pi},
\label{eq:TNDR}
\end{equation}
with
\begin{equation}
\tilde{T}_{DR}^{\rm on}(E)=\sum_{i,j=0}^N \lambda_{ij} (2 \mu E)^{i+j}.
\label{eq:TTDR}
\end{equation}
Therefore by appropriate choice of the $M_N$ coefficients
$\lambda_{ij}$ one can fit the first $M_N$ coefficients in the
effective range expansion for $\frac{1}{T_{DR}^{\rm on}(E)}$
with no constraints at all. So, the amplitude generated by DR
is apparently not subject to the constraint (\ref{eq:bound}). 

An alternative mathematically elegant way to discuss the
delta-function potential $V_3^{(0)}$ is in terms of the self-adjoint
extension of the free Hamiltonian on a space of smooth functions
compactly supported away from the origin~\cite{Al88}.  It has been
shown that obtaining $r_e >0$ with a delta-function potential
$V_3^{(n)}$ defined in ways analogous to this is impossible for a
Hamiltonian which is a Hermitian operator on a space with
positive-definite norm~\cite{Fe95}. This is to be expected from the
Wigner bound (\ref{eq:WB}), which was originally derived assuming only
unitarity and causality. Since using DR allows amplitudes with $r_e >
0$ we conclude that the results given by DR for delta-function
potentials such as $V_3^{(2)}$ do not correspond to a conventional
quantum mechanical calculation with truly zero-range potentials.

\section{Discussion}

\label{sec-drdiff}

We have found several quantum mechanical problems where DR and cutoff
regularization schemes give different physical results.  Our findings
raise two important questions which we will discuss in this section.
First, it is well known that for calculations in quantum field
theory which are carried out to a given order in the (renormalized)
coupling constant these two regularization schemes give equivalent
renormalized amplitudes~\cite{Co84}. What then are the features of
the non-perturbative calculations considered here which lead to results
at variance with the familiar perturbative case?  Second, since 
the potential (\ref{eq:V}) has no meaning before a regularization
scheme is chosen, and yet different regularization schemes lead
to different physical amplitudes, we ask in what sense the expression
(\ref{eq:V}) is to be understood.

In order to address the first issue it is useful to give a simple
illustrative example of perturbative regularization and
renormalization to contrast with the non-perturbative calculations of
this paper. Consider the perturbative treatment of the fermion
self-energy in a theory of fermions interacting with scalars via a
Yukawa coupling. The Lagrangian with physical masses and couplings
takes the form
\begin{equation}
{\cal L}=\frac{1}{2} (\partial_\mu \phi \, \partial^\mu \phi - m^2 \phi^2)
+ \bar{\psi} (i \!\not\!\partial - M) \psi - g \bar{\psi} \phi \psi.
\end{equation}

\begin{figure}[h]
   \vspace{0.5cm}
   \epsfysize=2cm
   \centerline{\epsffile{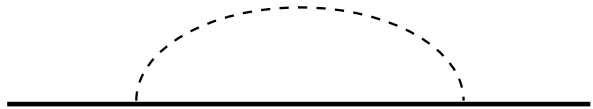}}
   \centerline{\parbox{11cm}{\caption{\label{fig2}
Fermion self-energy graph.
  }}}
\end{figure}

The graph shown in Fig.~\ref{fig2} corresponds to a fermion self-energy
\begin{equation}
\Sigma(p)=(-ig)^2 \int \frac{d^4k}{(2 \pi)^4} \frac{i}{\not\!p - \not\!k - M}
\frac{i}{k^2 - m^2}.
\end{equation}
This may be decomposed as $\Sigma(p)=A(p^2)\!\!\not\!p  + B(p^2)$ with
\begin{eqnarray}
A(p^2)&=&\int d^4k \, f(p^2,k^2,p \cdot k) - \frac{1}{p^2} 
\int d^4k \, f(p^2,k^2,p \cdot k) \, p \cdot k,\\
B(p^2)&=&M \int d^4k \, f(p^2,k^2,p \cdot k),
\end{eqnarray}
where
\begin{equation}
f(p^2,k^2,p \cdot k)=\frac{g^2}{(2\pi )^4}
\frac{1}{(p - k)^2 - M^2}\frac{1}{k^2 - m^2}.
\end{equation}
Observe that both $A$ and $B$ are divergent, and, in particular, that
the second integral in the expression for the coefficient function
$A(p^2)$ contains a linear divergence. 

In order to renormalize the free fermion Lagrangian we add the
counterterms
\begin{equation}
(Z_2 - 1) i \bar{\psi}\!\!\not\!\partial \psi - \bar{\psi} (Z_2 M_0 - M) \psi.
\end{equation}
The total fermion propagator, including $\Sigma(p)$, then becomes
\begin{equation}
d^{-1}(p)=Z_2 (\!\not\!p - M_0) - A(p^2) \!\!\not\!p  - B(p^2).
\end{equation}
The counterterms $Z_2$ and $M_0$ are then chosen so that
\begin{eqnarray}
M &=& Z_2 M_0 + B(M^2),\\
1 &=& Z_2 - A(M^2), \label{eq:Z2renorm}
\end{eqnarray}
thus giving a propagator with unit residue at the pole $\not\!p=M$. 

In DR the linear divergence in $A(M^2)$ is discarded, by prescription.
However, Eq.~(\ref{eq:Z2renorm}) indicates that the neglect of this
divergence merely results in an altered definition of the counterterm
$Z_2$. Since the bare Lagrangian is unobservable the physical results
obtained in DR and cutoff schemes will therefore be entirely
equivalent.  We have presented this example in some detail because it
is a paradigm for the way perturbative renormalization is carried out
in quantum field theory.  The key feature here is that the
renormalization is additive: the counterterms are linearly related to
the divergences, and only the sum of the two has physical meaning.
When this is the case a calculation done using cutoff regularization
may be mapped to one using DR by simply changing the definition of all
the counterterms. Under such a mapping the results for renormalized
amplitudes will remain the same. However, this means that DR {\it is
not solely} a regularization scheme.  DR regulates divergent
integrals, but also performs an implicit renormalization by setting
all power-law divergences to zero.  This automatic ``subtraction'' can
be an asset in perturbation theory because it provides an economical
way to ensure that symmetries like gauge invariance and chirality are
manifest throughout a loop calculation.

In contrast to the above example regularization and renormalization in
non-perturbative computations often have subtleties not present in the
perturbative case. These subtleties occur because the relation between
counterterms is more complicated than in perturbative quantum field
theory. The renormalization conditions often take on a form different
to that familiar from perturbative calculations, where a physical
observable is usually expressed as the sum of a divergence and a
counterterm.

Consider the case of the potential $V^{(0)}$ discussed in
Section~\ref{sec-dddf}. After iteration there is one divergence and
one counterterm ($1/C$), with the two linearly related.  In $d=2$ and
$d=3$ the divergence can be absorbed into the counterterm.  It is
therefore no surprise that DR and cutoff schemes give equivalent
results in these cases.  However, in $d\geq 4$ energy-dependent
divergences appear in the integral. These cannot be absorbed into the
counterterm. In DR there are no divergences in odd dimensions $d\geq
3$ and so DR and cutoff schemes are inequivalent in these dimensions.

The potential of Section~\ref{sec-df2} provides a good example of the
fundamental differences between perturbative and non-perturbative
renormalization. Three divergences, $I_1$, $I_3$, and $I_5$ appear in
the unrenormalized amplitude, and only two counterterms $C$ and $C_2$
are present. The interrelation of these divergences and counterterms
is highly nonlinear---a feature particular to non-perturbative
renormalization. Under such circumstances DR is not
necessarily equivalent to cutoff schemes.  DR's disregard of linear
divergences leads to a renormalized amplitude completely different to
that found in cutoff regularization. Furthermore, the relative size of
various divergences is meaningless in DR, and so the Wigner bound $r_e
\leq 0$---which in cutoff schemes arises as a consequence of the
non-linear relation of the various divergences---does not apply.

We turn now to our second issue: the interpretation of the potential
(\ref{eq:V}).  The potential defined by Eq.~(\ref{eq:V})
is intrinsically meaningless: it makes sense only after
regularization. Sections~\ref{sec-dddf} to
\ref{sec-wb} demonstrate that in certain circumstances the physical
amplitude which is obtained will be different in different
regularization schemes. Therefore, we must choose which regularization
scheme to use in order to give meaning to this otherwise nonsensical
potential. Such a choice can only be made by considering the context in which
such a delta-function potential might be used. We know of
only one circumstance in which a delta-function potential can be
profitably used. For this circumstance we will show that a cutoff scheme
is the appropriate form of regularization.

Consider a general non-local potential $V(r,r')$ which is of range
$R$.  Suppose that the structure of $V$ is tuned so that the first
$M_N$ terms (see Eq.~(\ref{eq:MN})) in the effective range expansion
are governed by a length scale $r_0 \gg R$. Despite this tuning, the
$N+1$th and higher terms in the effective range expansion will, in
general, be governed by the scale $R$, not the scale $r_0$. If we
attempt to model this potential by $V^{(N)}$ from Eq.~(\ref{eq:V})
then we know, from our previous arguments, that DR and cutoff schemes
give different results.  There are two arguments in favor of the use
of a cutoff scheme with the cutoff taken to infinity in this
particular problem.

First, we know from Wigner's bound (\ref{eq:WB}) that for $r_0 \gg R$
all coefficients in the effective range expansion beyond the
scattering length must be negative. As was shown in
Sections~\ref{sec-df2} and \ref{sec-wb} cutoff regularization (with
$\beta \rightarrow \infty$) reproduces this known feature of
scattering from $V$. By contrast, DR imposes no such restriction on
the effective range, shape parameter, etc.

Second, Eqs.~(\ref{eq:TNDR}) and (\ref{eq:TTDR}) show that DR predicts
that coefficients in the effective range expansion beyond those fit by
the renormalization conditions will be of scale $r_0$. By contrast, in
cutoff regularization one expects these higher-order effective range
expansion coefficients to be of scale $1/\beta$, where $\beta$ is the
cutoff. If we interpret $\beta \sim 1/R$ this accords with our
physical intuition about the size of these coefficients.  Thus, cutoff
regularization provides a physically intuitive procedure for giving
meaning to the potential (\ref{eq:V}). It just reintroduces the range
of the potential $V$ we are trying to model via a delta
function. Consequently, it will only be useful to take the limit
$\beta \rightarrow \infty$ is there is truly a wide separation of
scales in the problem, i.e.  $R \ll r_0$. This condition is not obeyed
in nucleon-nucleon scattering, except if only the scattering length is
fitted in the ${}^1S_0$ channel. Therefore, our arguments here do not
have direct relevance for effective field theory treatments of
nucleon-nucleon scattering. The implications of this work for that
problem will be elucidated in a forthcoming paper~\cite{Be97}.

This demonstrates that for this class of problem cutoff
regularization provides a controlled way of reintroducing the
breakdown of delta-function behavior which we know must occur at short
distances. On the other hand DR is an {\it ad hoc} modification of
short-distance behavior. We would argue that both in perturbation
theory and in a non-perturbative context one must verify that DR is a
sensible procedure by regularizing using a cutoff scheme and comparing
the calculated physical observables obtained via the two methods. Here
we have exhibited an instance in which cutoff regularization gives
results in accord with the behavior in an underlying theory while DR
does not. At least in this problem, DR treats short-distance physics
incorrectly.

\section*{Acknowledgments}
We thank U.~van Kolck for a number of useful discussions and for his
helpful remarks on the manuscript. We also thank S.~Weinberg for an
interesting conversation and I.~Afnan for his comments on the
manuscript. This research was supported in part by the
U. S. Department of Energy, Nuclear Physics Division (grant
DE-FG02-93ER-40762).

\end{document}